\documentclass[prl,showpacs,twocolumn]{revtex4-1} 
\usepackage{graphicx}
\usepackage{amsmath}
\usepackage{amssymb}
\usepackage{bm}

\newcommand{\be}{\begin{eqnarray}}
\newcommand{\ee}{\end{eqnarray}}
\newcommand{\nn}{\nonumber}
\newcommand{\bn}{\begin{enumerate}}
\newcommand{\en}{\end{enumerate}}

\def\IC{\mathbb{C}}

\def\IP{\mathbb{P}}

\def\CL{{\cal L}}

\def\CN{{\cal N}}

\def\CW{{\cal W}}

\def\CZ{{\cal Z}}

\def\a{\alpha}

\def\d{\delta}
\def\e{\epsilon}

\def\l{\lambda}

\def\s{\sigma}

\def\D{\Delta}

\def\L{\Lambda}

\def\goto{\rightarrow}

\def\det{{\rm det}}

\def\vol{\mbox{vol}}

\def\jmath{{j}}

\begin{document}

\title{A new integral formula for supersymmetric scattering amplitudes in three dimensions}
\author{Yu-tin \surname{Huang}$^a$ and Sangmin \surname{Lee}$^b$}
\affiliation{
${}^a$Department of Physics and Astronomy, UCLA, Los Angeles, CA 90095-1547, USA\\
${}^b$Department of Physics and Astronomy 
\& College of Liberal Studies \\ 
Seoul National University, Seoul 151-742, Korea}

\begin{abstract}
We propose a new integral formula for all tree-level 
scattering amplitudes of $\CN=6$ supersymmetric Chern-Simons theory. 
It resembles the Roiban-Spradlin-Volovich-Witten formula for $\CN=4$ supersymmetric Yang-Mills theory based on a twistor string theory formulation. Our formula implies that the $(2k)$-point tree-level amplitude is closely related to degree $(k-1)$ curves in $\IC\IP^{k-1}$.  

\end{abstract}

\pacs{11.15.Yc,11.30.Pb,11.55.Jy,02.30.Ik}
\keywords{S-matrix, Supersymmetry, Chern-Simons theory}

\maketitle


\paragraph{Introduction.} 

One of the major milestones in recent developments in the study of scattering amplitudes of gauge field theories was Witten's reformulation 
of the tree-level amplitudes of $\CN=4$ supersymmetric Yang-Mills theory (SYM$_4$) 
in the framework of a twistor string theory \cite{Witten:2003nn}. 
%
This formulation has inspired and influenced, directly or indirectly, a large number of subsequent developments; 
see {\it e.g.} \cite{review} for a review.

The twistor-string formulation led to the so-called connected prescription by Roiban, Spradlin and 
Volovich (RSVW) \cite{Roiban:2004yf}, which renders various properties among partial ordered tree-level amplitudes, such as the U(1) decoupling or Kleiss-Kuijf relations~\cite{Kleiss:1988ne}, either manifest or straightforward to prove.

On the other hand, it was shown by Arkani-Hamed, Cachazo, Cheung and Kaplan (ACCK) \cite{ArkaniHamed:2009dn} that, to all orders in perturbation theory, the leading singularities of SYM$_4$ are given by the residues of a contour integral over a Grassmannian manifold. This formulation manifests the SU(4$|$4) dual superconformal invariance~\cite{Drummond:2008vq}, or equivalently the Yangian invariance~\cite{Drummond:2009fd}, of individual amplitudes. 
Further investigation on the contour for tree-amplitudes, which are given by particular combinations of the leading singularities, led to the realization that ACCK formula evaluated for `tree contour' is equivalent to RSVW via a smooth deformation  \cite{Bullimore:2009cb,ArkaniHamed:2009dg, Bourjaily:2010kw}.

A natural question is whether these two formulas, and the special properties of SYM$_4$, can be generalized to theories with less supersymmetry or theories in 
other spacetime dimensions. A natural candidate would be the $\CN=6$ supersymmetric Chern-Simons theory (SCS$_6$, a.k.a. ABJM) \cite{Aharony:2008ug,Hosomichi:2008jd,Hosomichi:2008jb}, whose planar scattering amplitude was shown to also enjoy a hidden OSp(6$|$4) Yangian symmetry~\cite{Bargheer:2010hn,Huang:2010qy,Gang:2010gy}. 
Guided by how the Yangian symmetry is manifested in the ACCK formalism~\cite{Drummond:2010qh}, an analog of the ACCK formula 
for SCS$_6$ was proposed in \cite{Lee:2010du} and validated through explicit calculation. 
The similarity between planar scattering amplitudes of SYM$_4$ and SCS$_6$ leads one to consider the existence of a similar twistor string theory for SCS$_6$.  

The aim of this Letter is to show that, while a twistor string theory for SCS$_6$ has not been developed, an RSVW-like formula for tree amplitudes indeed exists. 
Our main strategy is to carry over the relation between ACCK and RSVW formulas 
\cite{Bullimore:2009cb,ArkaniHamed:2009dg, Bourjaily:2010kw} to three dimensions. 
We must emphasize that this is by no means, conceptually and technically, a trivial generalization. We find it likely that the existence of our new RSVW-like formula points toward a twistor-string-like reformulation of SCS$_6$. 

\paragraph{Review in four dimensions.}

In SYM$_4$, the ACCK formula states that the leading singularity of color-ordered 
amplitude for $n$ external particles with $k$ negative and $(n-k)$ positive helicity assignments can be written as
\be
\mathcal{L}_{n,k}(\CW) =
\int \frac{d^{k\times n}C}{\vol[{\rm GL}(k)]}
\frac{\prod_{m=1}^k\d^{4|4}(C_{mi}\CW_i)}{M_1(C) \cdots M_n(C)} \,.
\label{gra}
\ee
For each particle labelled by the index $i$, the super-twistor variable $\CW_i$ 
consists of four bosonic and four fermionic components and 
contains information on the external momentum and the super-multiplet structure. 
The integration variable $C$ is a $(k\times n)$ matrix. The $i$-th consecutive minor $M_i$ of $C$ is defined by $M_i(C) = \e^{m_1 \cdots m_k} C_{m_1 (i)} C_{m_2(i+1)} \cdots C_{m_k(i+k-1)}$. Geometrically, the 
space of $C$ modded out by the ${\rm GL}(k)$ `gauge' symmetry spans 
the Grassmannian coset space ${\rm Gr}(k,n) = {\rm U}(n)/\left({\rm U}(k)\times {\rm U}(n-k)\right)$. After gauge fixing and solving the bosonic delta functions, one has $(k-2)(n-k-2)$ integration variables left. Thus \eqref{gra} is understood as a $N\equiv(k-2)(n-k-2)$ dimensional contour integral, localized by the poles from the cyclic minors $M_i(C) $.  

It was shown in \cite{ArkaniHamed:2009dg} that the RSVW formula can also be written as an integral over a Grassmannian coset space:
\be
A_{n,k}(\CW) = 
\int 
\frac{d^{2\times n}\s}{\vol[{\rm GL}(2)]}
\frac{\prod_{m=1}^k\d^{4|4}(C_{mi}[\s]\CW_i)}{(1 2)(23) \cdots (n 1)} .
\label{twi}
\ee
Here, the integration variable $\s$ is a $(2 \times n)$ matrix 
and gets mapped to the $(k\times n)$ matrix $C[\s]$ by 
\be
\s = \begin{pmatrix}
a_1 \cdots a_n \\ b_1 \cdots b_n
\end{pmatrix}
\quad \rightarrow \quad
C_{mi}[\s] = a_i^{k-m}b_i^{m-1} \,.
\label{vero-map}
\ee
The bracket in \eqref{twi} is defined by $(i j)\equiv a_i b_j - a_j b_i$. Geometrically, the space of $\s$ modulo the ${\rm GL(2)}$ `gauge' symmetry 
spans ${\rm Gr}(2,n) = {\rm U}(n)/({\rm U}(2)\times {\rm U}(n-2))$. 

The map in (\ref{vero-map}) is known as the Veronese map which translate the  GL(2) transformation on the $\sigma$ variables to the GL$(k)$ transformation on the $C$ variables. 
If one intereprets ${\rm Gr}(k,n)$ projectively as a configuration of $n$ points in $\IC\IP^{k-1}$, \eqref{vero-map} implies that all $n$ points are localized on the image of a degree $(k-1)$ map from $\IC\IP^1$ to $\IC\IP^{k-1}$. 
To see how this localization is related to the original twistor string formulation, 
one takes a Fourier tranform from $\CW$ to its conjugate twistor variable $\CZ$, 
\be
\d^{4|4}(C_{mi}[\s]\CW_i) \goto
\int d z_m \prod_i \d^{4|4}(\CZ_i - z_m C[\s]_{mi}),
\label{fourier}
\ee
and observes that the tree-level amplitude is given by an integral over the moduli space of degree $(k-1)$ curves on which the twistor variables localize \cite{Witten:2003nn}. 




The localization can be conveniently achieved by a series of Veronese operators $S_l(C)$, with $l=1,\ldots,N$, which are quartic polynomials of (consecutive or non-consecutive) minors of $C$. 
By definition, the zero-locus of Veronese operators is preciely  
the image of the map \eqref{vero-map}. Thus 
evaluating the contour integral in ${\rm Gr}(k,n)$ on the poles of $S_l(C)^{-1}$ enforces that the ${\rm Gr}(k,n)$ variables can be put in the form of (\ref{vero-map}) up to a GL$(k)$ transformation.  
The final ingredient that links \eqref{gra} to \eqref{twi} was the realization \cite{Bullimore:2009cb,ArkaniHamed:2009dg, Bourjaily:2010kw} that \eqref{gra} can be written in a form such that the denominator of the integrand can be continuously deformed into products of $S_l(C)$. 
In particular, it was shown that the integrand in  \eqref{gra} can be written as: 
\be
\frac{1}{M_1(C) \cdots M_n(C)} 
=
\frac{H(C)}{S_1(C,t_1) \cdots S_N(C,t_N)}\bigg|_{t_l=0}\,, 
\label{vero-deform}
\ee
where $t_l$ is the deform parameter, and $S_l(C,t_l)$ becomes the Veronese operators when $t_l=1$. Thus by choosing the contour to be the zero-locus of $S_l(C)$ one finds an equivalence between \eqref{gra} evaluated on this `tree contour' to that of \eqref{twi}.


\paragraph{A new formula in three dimensions}

A Grassmannian-based formula for the leading singularities of SCS$_6$, 
analogous to the ACCK formula, was found in \cite{Lee:2010du}, 
\be
\CL_{2k}(\L) =\int \frac{d^{k\times 2k}C}{\vol[{\rm GL}(k)]}
\frac{\d(C_{mi} C_{ni})\d^{2k|3k}(C_{mi}\L_i)}{M_1(C) \cdots M_k(C)} \,.
\label{gra3}
\ee
The on-shell information of each external state are conveniently encoded by the three-dimensional `super-twistor' variables $\Lambda=(\l^{\a=1,2},\eta^{I=1,2,3})$. The on-shell multiplet are grouped into two super-fields with opposite (bose/fermi) statistics and adjacent multiplets in the color ordered amplitude must have opposite statistics.
By convention, we assign fermonic multiplets to odd-numbered sites. 
The alternating statistics is responsible 
for the sign factors in the cyclic symmetry,
\begin{align}
&A_{2k}(1,2,\ldots,2k-1,2k) 
\nn \\ 
&\qquad = (-1)^{k-1} A_{2k}(3,4,\ldots,2k-1,2k, 1,2) \,,
\label{cy3}
\end{align}
and the so-called $\L$-parity \cite{Gang:2010gy}, 
\be
A_{2k}(\ldots, -\L_i, \ldots) = (-1)^i A_{2k}(\ldots, \L_i, \ldots) \,.
\label{lp3}
\ee

The integration variable $C$ in \eqref{gra3} is a $(k\times 2k)$ matrix. 
The minors $M_i$ are defined in the same way as in \eqref{gra}.
A main novelty here, compared to \eqref{gra}, is the `orthogonal' constraint imposed by 
the delta function $\d(C_{mi}C_{ni})\equiv\d(O_{mn})$, 
which makes the domain of the integration to be an orthogonal Grassmannian OG$(k,2k)= {\rm O}(2k)/{\rm U}(k)$. After gauge fixing and solving the bosonic delta functions, one sees that \eqref{gra3} is in fact a $(k-2)(k-3)/2$ dimensional integration.

In this Letter, we propose a new (RSVW twistor-string-like) formula that gives the tree-level amplitudes of 
SCS$_6$:
\be
A_{2k}(\L) = \int\frac{d^{2\times n}\s}{\vol[{\rm GL}(2)]}
\frac{J \, \D \,\prod_{m=1}^k \d^{2|3}(C_{mi}[\s]\L_i)}{(1 2)(23) \cdots (n 1)}\,.
\label{twi3}
\ee
The domain of the integration and the product of consecutive $(2\times 2)$ 
minors in the integrand as well as the presence of the Veronese map are the same as in the four dimensional formula \eqref{twi}. 
In addition, it contains two factors, $J$ and $\D$, that have no counter-parts in four dimensions. 
The delta-function constraint,
\be
\Delta = \prod_{j=1}^{2k-1} \d\left(\sum_i a_i^{2k-1-j} b_i^{j-1}\right)\,,
\label{null-vero-const}
\ee
is inherited from $\d(O_{mn})$ in \eqref{gra3} (more on this below). 
The factor $J$ is a rational function of $\s_i$:
\begin{align}
&J= \frac{{\rm Num}}{{\rm Den}} \,, 
\;\;\;\;\;
{\rm Den} = \prod_{1\le i<j \le k} (2i-1 , 2j-1) \,,
\nn \\ 
& {\rm Num} = 
\det_{1\le i,j \le 2k-1} (a_i^{2k-1-j} b_i^{j-1}) 
\nn \\
&\quad\quad\; = \prod_{1\le i,j \le 2k-1} (i,j)
\label{JJ}
\end{align}
We first confirm the GL(2)=GL(1)$\times$SL(2) `gauge' symmetry. 
The SL(2) invariance 
is clear since the integrand is written entirely in terms of the brackets $(i,j)$. 
It is also straightforward to check the GL(1) weight cancels in \eqref{twi3}. 
The GL(2) gauge symmetry of the integral \eqref{twi3} makes the domain of the integration to be 
Gr$(2,2k)$ whose dimension is $(4k-4)$. This dimension counting will be crucial 
in the derivation of \eqref{twi3} from \eqref{gra3} to be discussed below. 
\paragraph{Consistency checks.}

We now show that the new formula \eqref{twi3} passes a few non-trivial 
tests. 
First, the $\L$-parity \eqref{lp3} simply follows from the delta function $\d^{2|3}(C_{mi}\L_i)$.

Second, consider the cyclic symmetry \eqref{cy3}. 
The numerator (Num) contains only $(2k-1)$ out of $(2k)$ possible columns, 
so it may appear to spoil the cyclic symmetry.  
But, the constraint \eqref{null-vero-const} states that 
the sum of all $(2k)$ columns vanishes, i.e., they are linearly dependent and hence (Num) is cyclic invariant. 
Under the cyclic shift by two sites, the denominator (Den) 
merely reshuffles the same $(2i-1,2j-1)$ factors in a different ordering, 
and comes back to itself 
with the desired $(-1)^{k-1}$ factor. 


Finally, we consider the three-algebra based \cite{Bagger:2008se} Kleiss-Kuijf (KK) type relations  \cite{Kleiss:1988ne} which has not been discussed in the literature: 
\be
\sum_P (-1)^P A_{2k}(1,P(2),\ldots,2k-1,P(2k)) = 0\,,
\label{kkid}
\ee
where $P$ denotes any permutation of the even labels $\{2,4,\ldots,2k\}$. This identity can be deduced by representing the color dressed amplitude in terms of the three-algebra color factors, which are products of the four-indexed structure constant $f^{abcd}$ of the three-algebra. Converting to the trace basis leads to the linear relations in \eqref{kkid} between the color ordered amplitudes. Alternatively, the relation can be proven by induction using recursion relations~\cite{Gang:2010gy}. In \eqref{twi3}, only the cyclic denominator factor is relevant for \eqref{kkid}, as all other parts are invariant under permutation of the even sites. One can then easily check that the factor
\be
\frac{1}{(1 2)(2 3)\cdots (2k-1,2k)(2k,1)} 
\ee
satisfies \eqref{kkid}. As this factor is also present in \eqref{twi}, one immediately concludes that Yang-Mills amplitude also satisfies \eqref{kkid}. This is not surprising given the fact that for even point amplitudes, once can express the color factors of Yang-Mills in terms of $f^{abcd}=f^{ab}\,_ef^{cde}$.

Unlike the $\Lambda$-parity and cyclic symmetry, which is built into \eqref{gra3} and hence valid for all residues of the Grassmannian, the KK identity \eqref{kkid} is only valid for the tree amplitudes. Therefore, the fact that \eqref{twi3} does satisfy this identity to all orders is a non-trivial support that this is indeed equivalent to the Grassmannian \eqref{gra3} evaluated along the correct `tree contour'. 

\paragraph{Constructive proof up to eight points.}
We have succeeded in deriving \eqref{twi3} from \eqref{gra3}
by explicit computations up to $k=4$. 
The presence of the Veronese map in \eqref{twi3}, indicates that it must be related to \eqref{gra3} evaluated on a contour that is on the zero-locus of the Veronses operators. For $k=2$, the map \eqref{vero-map} is an identity map and the equivalence is straightforward. 

For $k=3$, \eqref{gra3} is a zero dimensional integral and there are no contours to choose. This implies that at $k=3$ \eqref{gra3} must be equivalent to \eqref{twi3} and the six-point Veronese operator \cite{ArkaniHamed:2009dg} must have trivially vanished. 
As we will see, the vanishing of the Veronese operator is in fact implied by the orthogonal constraint. For $k=3$, after gauge fixing but prior to solving any bosonic delta functions, we have 9 integration variables in \eqref{gra3} and 8 variables in \eqref{twi3}. We are free to introduce any coordinate in the direction normal to $\s$ as long as the resulting Jacobian is non-singular, 
\be
d^9C|_{\rm g.f.} = \det\left(\frac{\partial C}{\partial \s},\frac{\partial C}{\partial \tau}\right) d^{8}\s|_{\rm g.f.} d\tau \,.
\label{jac3}
\ee
To integrate out $\tau$, we note the following general fact. 
The components of the matrix $O_{mn} \equiv C_{mi} C_{ni}$ 
with the same $(m+n)$ take the same value on the image of the map \eqref{vero-map}. Thus on the image of the map \eqref{vero-map}, some of the orthogonal constraints become redundant. In particular, for $k=3$ with a convenient choice of the $\tau$ coordinate, $C = C[\s] + \tau \d_{m1}\d_{i1}$, one can write
\begin{eqnarray}
\d\left(O_{22}\right)\d\left(O_{13}\right)
=\d\left(O[\s]_{22}\right)\d\left(\tau b^2_1\right) \,.
\end{eqnarray}
Thus we see that, at six-point, the orthogonal constraint indeed localizes the $C_{mi}$ onto the image of the Veronese map and hence the vanishing of the six-point Veronese operator.
Integrating out $\tau$ and using some identities implied by the remaining parts of $\d(O_{mn})$, 
we find that \eqref{twi3} indeed follows form \eqref{gra3}. 

The generic situation begins at $k=4$ where \eqref{gra3} is a one dimensional contour integral. There are four distinct Veronese operators for $k=4$, and hence four $\tau$ coordinates, which implies that the orthogonal constraint must trivialize three of them. 
From $\d(O_{mn})$ we see three pairs of degenerate constraints on the image of the map \eqref{vero-map}; $(O_{13}, O_{22})$, $(O_{14}, O_{23})$ and $(O_{24}, O_{33})$. With suitable choice of $\tau$, one obtains three constraints linear in $\tau$'s. 
The last constraint must appear as poles from the presence of the remaining Veronese operator in the denominator of \eqref{gra3} 
in a way analogous to the four dimensional case we reviewed earlier.  

There are two hints we can use to determine the wanted factor 
from the minors. One hint is that when evaluating the integral 
\eqref{gra3}, the tree-amplitude is given by the sum of either poles in $1/(M_2M_4)$ \cite{Gang:2010gy}. The other hint comes from 
the known form of $S_l(C)$ in \eqref{vero-deform} for $n=8$, $k=4$ \cite{Bourjaily:2010kw}. 
By combining the two pieces of information, we are led to deform 
the integrand as follows.
\begin{align}
&\frac{1}{M_1 M_2 M_3 M_4} 
\; 
\goto 
\;
\frac{H}{S} \,,
\quad 
H = \frac{(4631)(4712)}{M_1 M_3} \,, 
\nn \\
&S = M_2 M_4 (4631)(4712) - M_1 M_3 (4167)(4275)\,,
\end{align}
where $(ijkl)$ denotes the minor composed of the four non-consecutive columns of $C$. 
%
%
Now, introducing a suitable set of coordinates $\tau_l$ $(l=1,2,3,4)$ 
normal to the $\s$ directions 
and integrating them out using the four factors, 
we find that \eqref{twi3} indeed follows from \eqref{gra3} evaluated on the tree contour.

\paragraph{A discussion on general $k$.}


For general $k$, it is always possible to introduce $\tau$ coordinates 
in the small neighborhood of the image of the map \eqref{vero-map}. 
The total number of $\tau$ coordinates, or equivalently Veronese operators, is 
\be
\dim[{\rm Gr}(k,2k)] - \dim[{\rm Gr}(2,2k)] = (k-2)^2 \,.
\ee
As we noted earlier, the components of $O_{mn}$ with 
the same $(m+n)$ take the same value, so the differences 
between them can be used to integrate out $\tau$'s. 
The number of such differences is $(k-2)(k-1)/2$. 
%
%
%
%
%
%
The remaining $(k-2)(k-3)/2$ $\tau$'s should be 
integrated out by a contour integral around `poles' from 
a suitably deformed minors as was the case in four dimensions. 
In other words, 
\be
(k-2)^2 = \underbrace{\frac{(k-2)(k-3)}{2}}_{\rm minor} 
+ \underbrace{\frac{(k-2)(k-1)}{2}}_{\d(CC^T)} \,.
\ee
The number of poles needed in the minors to integrate out $\tau$'s 
coincides exactly with the number of poles needed for the 
full evaluation of the Grassmannian formula \eqref{gra3} 
as discussed in \cite{Lee:2010du,Gang:2010gy}. 
This matching indicates that the problem of deforming the 
minors to integrate out $\tau$'s is closely related to 
classifying the independent Veronese operators under the orthogonal constraint.




%

\paragraph{Discussion.}
So far we have shown up to $k=4$ that the `tree contour' of \eqref{gra3} is on the zero-locus of the non-trivial Veronese operators, leading to the integral representation of tree-level amplitudes in \eqref{twi3}. While it is unclear whether for higher $k$ the `tree contour' should continue to be on the zero-locus of the Veronese operators, leading to the validity of \eqref{twi3}, the fact that \eqref{twi3} renders tree-level amplitude relations such as the KK relations \eqref{kkid} straightforward supports the assertion and the validity of the formula. Furthermore, the fact that the orthogonal constraint enforces some of the Veronese to vanish also supports our identification of the tree-contour in \eqref{gra3}. It would be interesting if one can further test the formula by extracting its double soft behavior, as well as possible non-trivial amplitude identities that arise from color kinematic dualities \cite{Bargheer:2012gv} similar to that found for Yang-Mills \cite{Bern:2008qj}. 

Finally, the fact that the tree amplitude corresponds to localization of the orthogonal Grassmanian \eqref{gra3} on degree $(k-1)$ curves is highly suggestive of a novel twistor string formulation. For instance, a Fourier transform to a `dual' space, in a way analogous to \eqref{fourier},
shows that the amplitude in the dual space is indeed localized on degree $(k-1)$ curves. However, a crucial difference from SYM$_4$ is the additional orthogonal constraint. It would be interesting to understand the geometric picture of such constraint. 
 




\paragraph{Acknowledgments.}
We thank the Institute for 
Advanced Study at Princeton for hospitality 
where this work was initiated. 
We are especially grateful to N. Arkani-Hamed 
for several inspiring discussions. 
We also thank D. Gang and E. Koh for collaboration on 
an attempt for a recursive proof, 
and 
S. Caron-Huot, J. Maldacena, C. S\"amann, A. Volovich and E. Witten for 
stimulating conversations.   
%
%
The work of YTH was supported by the US Department of Energy under 
contract DE–FG03– 91ER40662.
The work of SL was supported in part by National 
Research Foundation of Korea (NRF) Grants  
2009-0072755, 2009-0084601 and 2012R1A1B3001085.

\end{document}